# Atomistic Investigation of Elementary Dislocation Properties Influencing Mechanical Behaviour of $Cr_{15}Fe_{46}Mn_{17}Ni_{22}$ alloy and $Cr_{20}Fe_{70}Ni_{10}$ alloy

Ayobami Daramola[a*], Anna Fraczkiewicz[a], Giovanni Bonny[b], Akiyoshi Nomoto[c], Gilles Adjanor[d], Christophe Domain[d], Ghiath Monnet[d].

[a]MINES Saint-Etienne, Université de Lyon, CNRS, UMR 5307 LGF, Centre SMS, 42023, ´ Saint-Etienne, France

[b]SCK CEN, Nuclear Materials Science Institute, Boeretang 200, B-2400 Mol, Belgium

[c]Central Research Institute of Electric Power Industry, 2-6-1 Nagasaka, Yokosuka-shi, Kanagawa 240-0196, Japan

[d]EDF Lab, Département Matériaux et Mécanique des Composants, Les Renardières, F-77250 Moret sur Loing, France

Corresponding author: Ayobami Daramola[a], Ghiath Monnet[d]
Email address: ayobami.daramola@emse.fr ghiath.monnet@edf.fr

**Abstract**

In this work, molecular dynamics (MD) simulations were used to investigate elementary dislocation properties in a Co-free high entropy (HEA) model alloy ($Cr_{15}Fe_{46}Mn_{17}Ni_{22}$ at. %) in comparison with a model alloy representative of Austenitic Stainless Steel (ASS) ($Cr_{20}Fe_{70}Ni_{10}$ at. %). Recently developed embedded-atom method (EAM) potentials were used to describe the atomic interactions in the alloys. Molecular Statics (MS) calculations were used to study the dislocation properties in terms of local stacking fault energy (SFE), dissociation distance while MD was used to investigate the dissociation distance under applied shear stress as a function of temperature and strain rate. It was shown that higher critical stress is required to move dislocations in the HEA alloy compared with the ASS model alloy. The theoretical investigation of simulation results of the dislocation mobility shows that a simple constitutive mobility law allows to predict dislocation velocity in both alloys over three orders of magnitude, covering the phonon drag regime and the thermally activated regime induced by dislocation unpinning from local hard configurations.

# 1 Introduction

Recent works have shown that some single-phase High Entropy Alloys (HEA) exhibit excellent mechanical properties, radiation-resistance properties and phase stability making them potential candidates for structural applications [1-10]. However, many authors have traced the origin of these properties to the specific behaviour of dislocations [11-28]. Face-Centered Cubic (FCC) HEA, such as Cantor alloy consisting of equimolar CoCrFeMnNi composition exhibits excellent mechanical properties and improved irradiation resistance [1]. On the other hand, the presence of Cobalt excludes Cantor alloy from structural applications in nuclear reactors, due to the production of [60]Co radioisotope in in-service conditions [19, 29-30]. This is one of the reasons for the design of new Co-free HEA alloys, mainly from the CrFeMnNi family [19, 29-35].

Currently, in the major types of nuclear reactor, the in-core components consist mostly of austenitic stainless steels (ASS) [36-38] and some authors [29, 36-37] have shown that ASS can be improved to exhibit better radiation damage resistance. Few authors like Kumar et al., [29], Gao et al., [35] have shown that Co-free alloy can be used for structural application to overcome this extreme environment and implemented for the next-generation nuclear reactors.

Irradiation hardening and embrittlement are key issues for structure alloys for applications under extreme conditions [29, 35, 39-43]. The irradiation hardening is associated to an increase in the critical stress of the material while the embrittlement is detected by the change of fractures properties. These issues are well investigated in ASS materials from the perspective of both experimental and numerical studies [43-45]. Findings from recent works like Zhao et al., [14, 15, 17] have shown that reduction of irradiation hardening is possible using FCC multi-principal element alloys (MPEAs) like HEA. The authors explained that the fluctuation of SFE improves irradiation tolerance thereby limiting the number of defects and cluster growth under the cascade during neutron irradiation. Gao et al., [35] have also investigated the role of local SFE on the formation of defects in $Cr_{15}Fe_{46}Mn_{17}Ni_{22}$ and $Cr_{16}Fe_{37}Mn_{13}Ni_{34}$ (at. %) under irradiation conditions. The reports from these works [15, 17] suggest that high fluctuation of SFEs in MPEA would increase the critical stress for dislocation glide and therefore, improve the mechanical behaviour of the material.

To the knowledge of the authors, there is no detailed work that would investigate the elementary dislocation behaviour in FCC CrFeMnNi alloys, especially their dissociation and mobility under different loading conditions. Experimentally, dislocation mobility is difficult to characterize and therefore molecular dynamics (MD) is often adopted to overcome this problem. Over the years, MD has been used to investigate the dislocation behaviour of materials of different crystallographic structures: FCC [14-17, 27, 43, 46, 47], Body-Centered Cubic (BCC) [48-50] as well as Hexagonal Close-Packed (HCP)[51, 52].

In our previous work [46], we developed a new embedded atom method (EAM) type interatomic potential to describe FCC CrFeMnNi systems. This is one of the few available EAM potentials that have been able to describe the dislocation properties of CrFeMnNi systems and their subsystems. This potential was fitted to reproduce good agreement for elastic constants, stacking fault energy and phase stability compared to experimental measurement, density functional theory and thermodynamic modelling. Gao et al., [35] recently used this potential to describe the effect of local stacking fault energies in Co-free alloys. To validate and confirm our results, some of the predictions of this potential are compared with those of a ternary potential [47] dedicated to the CrFeNi system. In the following, this potential is referred to as EAM-11, while the potential in [46] is called EAM-21. The default potential to be considered when it is not specified is EAM-21. When EAM-11 is used, it will be explicitly mentioned. The objective of this work is to investigate and compare elementary properties affecting the dislocation behaviour such as the fluctuation of the local SFE, and dislocation dissociation and mobility in the ASS and the HEA model alloys. A constitutive description of the dislocation mobility in the two systems is reported, providing insight into the mechanical characteristics of these two model alloys.

## 2 Methods

### 2.1 *Selection of model alloys*

This present study is general and applicable to any austenitic stainless steel CrFeNi, especially 300 series of steel of American standard AlSl (304 and 316-type) and Co-free HEAs such as FCC CrFeMnNi [19, 29-35]. The calculations are narrowed to $Cr_{20}Fe_{70}Ni_{10}$ at. % and $Cr_{15}Fe_{46}Mn_{17}Ni_{22}$ at. %. The reasons for the selection of these alloys are as follows:

a) Olszewska [33] work has reported that $Cr_{15}Fe_{46}Mn_{17}Ni_{22}$ alloy also called "Y3 grade" exhibit properties that are very similar to the Cantor alloy [1]. The interatomic potential we recently developed [46] fits the SFE of $Cr_{15}Fe_{46}Mn_{17}Ni_{22}$ to be 26 mJ/m$^2$ so it can follow the calculations of SFE by Zaddach et al., [53] and Smith et al., [13] for Cantor alloy which give values in the range of 26-28 mJ/m$^2$. The SFE is important to reproduce a reliable dislocation behaviour and hence, one of the reasons for $Cr_{15}Fe_{46}Mn_{17}Ni_{22}$ alloy selection.

b) In a previous work carried out by Kumar et al., [29], $Cr_{18}$(FeMnNi) wt. % alloy was shown to have a better radiation resistance than CrFeNi austenitic alloys. However, no work has given an in-depth understanding of irradiation conditions for $Cr_{15}Fe_{46}Mn_{17}Ni_{22}$ excepted Gao et al., [35] that experimentally investigated the ratio of faulted Frank loop against perfect loop and discovered that the formation of the defects depends on the fluctuations of SFE. Yet, no information was reported on the plasticity of $Cr_{15}Fe_{46}Mn_{17}Ni_{22}$, especially on the collective behaviour of defect character either edge or screw dislocation.

## 2.2 Computational methods

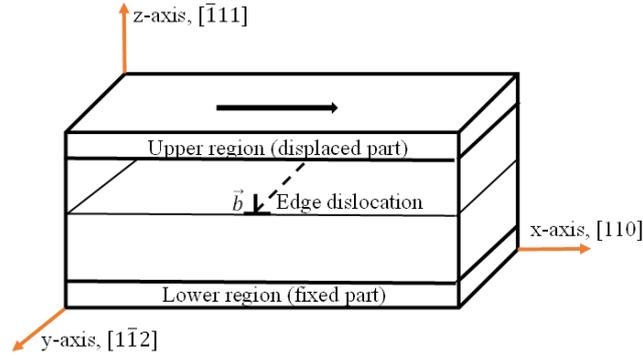

Figure 1: Coordinate system and geometry of the calculation box used for the present study. The periodic boundary condition is along with *x* and *y* directions, while the free boundary condition is along the z-direction. An edge dislocation is inserted in the middle by joining two half crystals and stress is induced by deformation rate as explained in the text.

EAM-11 potential [46] and EAM-21 potential [47] was used to calculate the dissociation distance and motion of dislocations for $Cr_{20}Fe_{70}Ni_{10}$ and $Cr_{15}Fe_{46}Mn_{17}Ni_{22}$ alloy respectively. The calculations based on EAM-11 potential was used in a classical MD code developed in the Tokyo University of Science (MicroMD). EAM-21 potential was used in two different MD codes in this work: DYMOKA code [54] for molecular statics (MS) and Large-scale Atomic/Molecular Massively Parallel Simulator (LAMMPS) software [55] for molecular dynamics (MD) simulations. When the box dimensions are too low, say few nanometers, the partial dislocations strongly interact with their images through the periodic boundary and with the fixed atomic layers. In this case, the box dimensions affect all the observed dislocation properties. As shown in figure 1, the simulation boxes dimensions are $45 \times 30 \times 20 \: nm^3$ in [110], [1$\bar{1}$2] and [$\bar{1}$11] directions, respectively, which correspond to about 2.59 million atoms in the cell. These dimensions are large enough to avoid spurious effects induced by the simulation box size. The model material is relaxed and then thermalized at given temperatures. The periodic boundary conditions are applied on the (110) and (1$\bar{1}$2) planes. For the dislocation motion calculations, two different methods were used to introduce the ½ <110>{111} edge dislocation in the simulation box. The method called $d_+$ was initially reported by Osetsky et al [56] in which the dislocation is initially spread over the whole simulation width. The second one is called $d_-$ method was described by Rodney et al. [57] and corresponds to the generation of a compact dislocation (Volterra dislocation). The atoms on the top and bottom surfaces are fixed and then displaced in the [110] direction at a constant displacement rate to apply shear deformation at a given strain rate. Since the temperature change during deformation is typically only several kelvins, no temperature control was performed during the deformation calculations. The shear stress is calculated from the component of the force on the fixed atoms parallel to the Burgers vector. The dislocation lines were identified by the common neighbour analysis (CNA) method [58] and the dislocation extraction algorithm (DXA) function [59].

The SFE calculations were carried out using the same fcc crystal simulation cell box and oriented directions. To determine the SFE, the simulation box sizes of about $10 \times 5 \times 10 \times 12$ containing around 6000 randomly distributed atoms and to analyze the role of local SFE, 6000 calculations with different random seeds within the calculation box were performed. The SFE is defined as:

$$\gamma_{SFE} = \frac{E_f - E_i}{A} \quad (1)$$

where $E_f$ and $E_i$ are the energies of the system after and before creating stacking fault, respectively, and $A$ is the area of stacking fault.

## 3 Results and discussion

### 3.1 *Elastic deformation response of the alloys*

The shear modulus on the {111} planes in the <110> directions of the $Cr_{20}Fe_{70}Ni_{10}$ and $Cr_{15}Fe_{46}Mn_{17}Ni_{22}$ alloy was investigated. The shear modulus is obtained by loading the simulation box (Fig. 1) and it equals the slope of the shear stress – shear strain curve when the dislocation does not move. Figure 2 shows the obtained response of the box at 300 K with a strain rate of $10^6\ s^{-1}$. Every peak of the curves corresponds to a rearrangement of the dislocation cores. Between the peaks the dislocation is immobile, and the slope can be unambiguously calculated. From our MD calculation, the shear modulus is estimated to be 59 GPa and 79 GPa for the ASS $Cr_{20}Fe_{70}Ni_{10}$ and HEA $Cr_{15}Fe_{46}Mn_{17}Ni_{22}$ alloys respectively. These values are compared with the values predicted by the anisotropic-elastic theory [60, 61] of the shear modulus on the {111} planes in the <110> directions. In cubic symmetry, it is given by $\mu = (C_{11} - C_{12} + C_{44})/3$. Table 1 shows the calculated values for both alloys. The latter is in close agreement with the values obtained from MD simulations and The difference in value is due to anisotropic-elastic theory's lack of a lattice force.

Table 1. The table contains the elastic constants predicted by EAM-21 potential, the estimated shear modulus from molecular dynamics calculations, anisotropy elastic theory denoted with $\mu^{MD}$ and $\mu^{ani-ela.}$ respectively (in GPa) for $Cr_{20}Fe_{70}Ni_{10}$ and $Cr_{15}Fe_{46}Mn_{17}Ni_{22}$ at 300 K.

| Alloy | $C_{11}$ (GPa) | $C_{12}$ (GPa) | $C_{44}$ (GPa) | $\mu^{MD}$ (GPa) | $\mu^{Ani-ela.}$ (GPa) |
|---|---|---|---|---|---|
| $Cr_{20}Fe_{70}Ni_{10}$ | 192 | 135 | 122 | 59 | 60 |
| $Cr_{15}Fe_{46}Mn_{17}Ni_{22}$ | 235 | 144 | 125 | 79 | 72 |

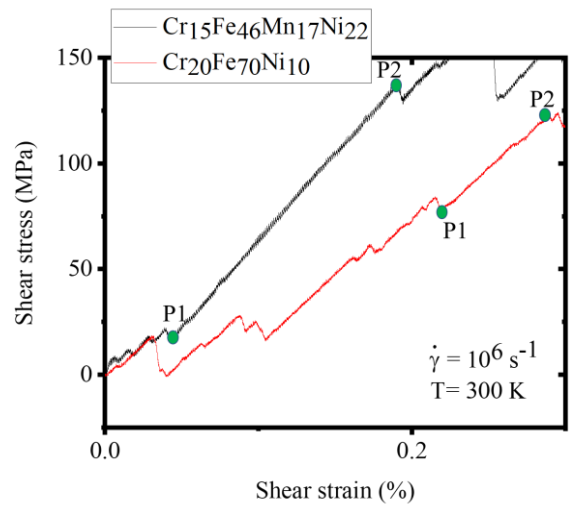

Figure 2: Stress-stress response to shear loading of the simulation box at 300 K ($\dot{\gamma} = 10^6\ s^{-1}$). The shear modulus is estimated between the peaks on each curve. P1 and P2 are the points to determine the slope of the line.

Experimentally, the shear modulus decreases with temperature [62, 63]. This agrees with the results presented in figure 3, in which we depict the evolution of the three elastic constants with temperature. All constants decrease significantly with temperature between 0 and 900 K. This qualitatively agrees with Haglund et al., [63] experimental explanation for HEA.

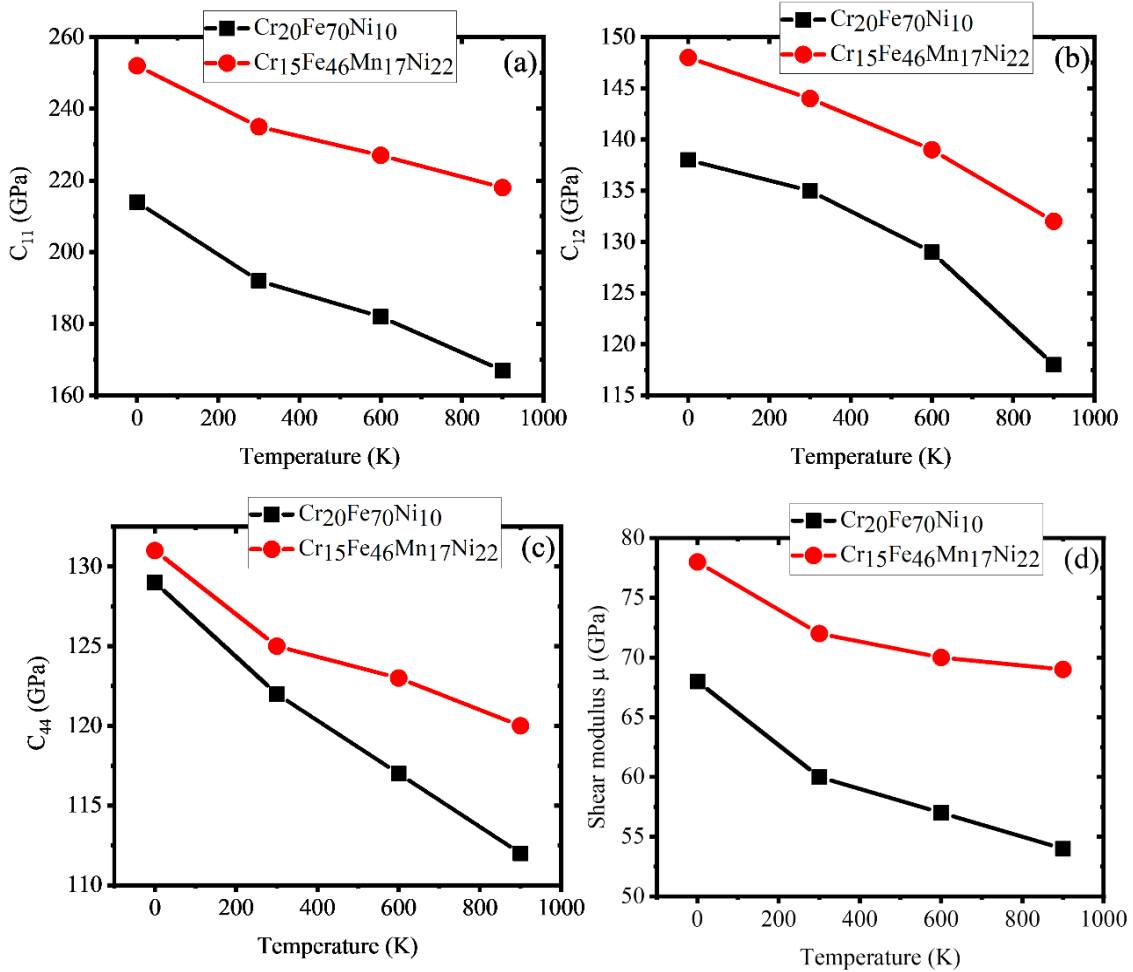

Figure 3: Temperature-dependence of the single elastic constant and the shear modulus $\mu$ for the austenitic stainless steel, $Cr_{20}Fe_{70}Ni_{10}$ and HEA, $Cr_{15}Fe_{46}Mn_{17}Ni_{22}$. (a-c) $C_{11}, C_{12}, C_{44}$ and (d) shear modulus $\mu$ for 0-900 K (The approach used can be seen in the appendix).

## 3.2 Investigation of the stacking fault energy

The <110>{111} edge dislocation in $Cr_{20}Fe_{70}Ni_{10}$ and $Cr_{15}Fe_{46}Mn_{17}Ni_{22}$ alloy dissociates into two Shockley partial dislocations separated by a stacking fault ribbon [64]. This follows the Shockley schema as:

$$\frac{a}{2}[110] \leftrightarrow \frac{a}{6}[121] + \frac{a}{6}[21\bar{1}] \qquad (2)$$

The atom distributions in the stacking fault ribbon region play an important role in the fluctuations of the dissociation distance. Moreover, both alloys have random atoms in nature Here, the effect of local atom distributions on the SFE was investigated. As earlier stated in section 2.2, the role of the local SFE distributions can be carried out by varying the element arrangement using different random seeds in the simulation as implemented in [14-18]. The SFE distribution is estimated with a Gaussian distribution given as:

$$g(x) = \frac{1}{\sigma\sqrt{2\pi}} exp^{-\frac{(x-\mu)^2}{2\sigma^2}} \qquad (3)$$

where the values of $\mu$, and $\sigma$ is the mean of the SFE and the standard deviation of the SFE distribution, respectively.

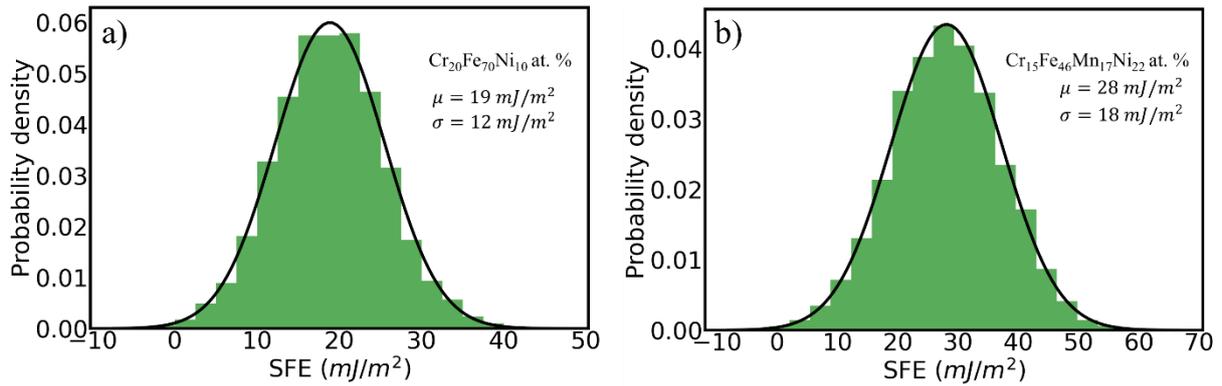

Figure 4: Distribution of stacking fault energy of a) $Cr_{20}Fe_{70}Ni_{10}$ and b) $Cr_{15}Fe_{46}Mn_{17}Ni_{22}$ alloy with 6000 random realizations. The mean and standard deviation are represented by $\mu$, and $\sigma$, respectively.

In figure 4, the distribution of the SFE of $Cr_{20}Fe_{70}Ni_{10}$ and $Cr_{15}Fe_{46}Mn_{17}Ni_{22}$ alloy is presented. The mean value of SFE of $Cr_{20}Fe_{70}Ni_{10}$ is estimated to 19 mJ/m$^2$ and $Cr_{15}Fe_{46}Mn_{17}Ni_{22}$ alloy is evaluated to be 28 mJ/m$^2$. However, the standard deviation which defines the fluctuations of the SFE distribution is approximately 12 mJ/m$^2$ for $Cr_{20}Fe_{70}Ni_{10}$ and 18 mJ/m$^2$ for $Cr_{15}Fe_{46}Mn_{17}Ni_{22}$ alloy. The calculations in figure 4 suggest that the fluctuation of the SFE distribution increases as the chemical composition becomes complex.

### 3.3 *Dislocation dissociation in the different alloys*

Since two opposite methods were used to introduce the edge dislocation in the simulation box, it is important to check their possible effects on the dislocation behaviour. To illustrate the effect of the selected method on the resulting dissociation, we apply both methods on the ASS ($Cr_{20}Fe_{70}Ni_{10}$) alloy using EAM-11 potential and analyze the obtained dislocation structure, relaxed at 300K (see Fig. 5).

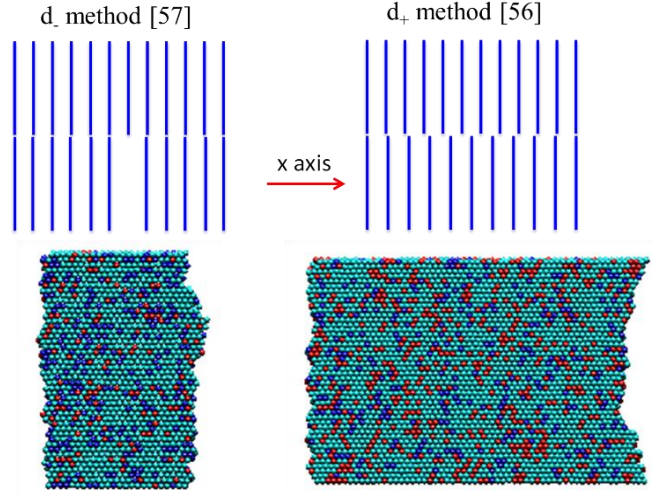

Figure 5: a sketch of the edge dislocation construction methods in the xz plane and the obtained dislocations after relaxation at 300 K in the xy plane.

The resulting dissociation is completely altered by the construction method. Is this discrepancy a simulation spurious effect or the consequence of a hidden physical mechanism? To answer this question, we follow the analysis reported in [64]. In pure FCC metals, the friction stress (solid solution) is negligible. The dislocation dissociation reaches equilibrium after relaxation in both methods. But in concentrated alloys, friction stress cannot be ignored, which affects simulation results as follows. When the edge dislocation in Fig. 1 is dissociated into two partials ("Partial_1" to the right and "Partial_2" to the left in the figure), every partial is submitted to the following force per unit length components $F_{app}$, $F_{int}$, $F_f$ and $F_\gamma$ corresponding respectively to the applied (Peach-Koehler) force induced by the applied shear stress $\tau_{app}$ ($F_{app}= \tau_{app}b$), the elastic repulsion between the partials, the friction (solid solution) force and the attraction force induced by the positive stacking fault energy (SFE) ($F_\gamma = \gamma$ per unit length). When forces are taken in absolute values, the total force on Partial_1 and Partial_2 are:

$F_1 = F_{int} - \gamma \pm F_f + \tau_{app} b /2$      (4)

$F_2 = - F_{int} + \gamma \pm F_f + \tau_{app} b /2$      (5)

where $b$ is the norm of the Burgers vector. The division by 2 of the applied stress term results from the projection of the partial Burgers vectors on the x –axis. We will see that these equations can explain all simulation results on the dislocation dissociation after relaxation.

Let us first define what we call the equilibrium dissociation width $d_o$. By definition, it corresponds to the dissociation width under zero applied stress and zero friction stress. Imposing $F_1 = F_2 = 0$ leads to $F_{int} = F_\gamma$ for the two partials. Recalling that for edge dislocations [64]:

$$F_{int} = \frac{Gb^2(2+\nu)}{24\pi(1-\nu)d}$$      (6)

Then, we obtain:

$$d_o = \frac{Gb^2(2+\nu)}{24\pi(1-\nu)\gamma}$$      (7)

In eqs. 4 and 5, the "±" sign accounts for the indetermination in the sign of the friction stress since it always opposes the dislocation motion. During relaxation of the initial atomic configuration in the $d_+$ method [56], the dislocation is initially dissociated over the whole simulation box width as shown in fig. 4. $d$ is thus larger than $d_o$ and partials tend to move towards each other. While in the $d_-$ method [57], the initial dislocation is compact and the two partials too close ($d \ll d_o$). Under the effect of $F_{int}$ partials tend to move away from each other. Consequently, during relaxation in the $d_-$ method, we can set the force balances as: $F_1 = F_{int} - \gamma - F_f$ and $F_2 = -F_{int} + \gamma + F_f$, while in the $d_+$ method, we have $F_1 = F_{int} - \gamma + F_f$ and $F_2 = -F_{int} + \gamma - F_f$. Therefore, in the $d_-$ method, the partials stop spreading when $F_{int} = \gamma + F_f$. The final dissociation width in the $d_-$ method is thus:

$$d_- = \frac{Gb^2(2+\nu)}{24\pi(1-\nu)(\gamma + F_f)} \qquad (8)$$

The spreading of the partials decreases with the friction stress and remains always lower than the equilibrium width $d_o$. The situation is different in the $d_+$ method. During relaxation, the only driving force for motion is the stacking fault energy $\gamma$, which must overcome the elastic repulsion and the friction stress. In order to have an idea of the order of magnitude of the different components, consider a separation of $d = 20$ nm, with a shear modulus of 70 GPa and Poisson's ratio of 0.3, one obtains from eq. 6: $F_{int} \approx 9.5$ mJ, while the average SFE is close to 28 mJ/m$^2$. Converting forces per unit length to shear stresses (using the relation $\tau = F/b$), we obtain $\tau_{int} \approx 38$ MPa and $\tau_\gamma \approx 100$ MPa. Consequently, in the $d_+$ method the partials will not be able to move towards each other if the friction stress is larger than $100 - 38 \approx 62$ MPa. As it will be seen in the next section, this is almost always the case. In the $d_+$ method, the obtained dissociation width is large, constant and temperature independent.

Now let us study the dissociation width under a given positive applied stress, large enough to induce dislocation motion. When the partials start moving in the positive x-direction, the friction stress on both partials in the two configurations ($d_-$ and $d_+$ methods) becomes negative (negative sign in equations 4 and 5). When the velocity is constant, the resultant force on both partials remains zero: $F_1 = F_2 = 0$. The applied stress balances completely the friction stress: $\tau_{app} b /2 - F_f = 0$. This is equivalent to the condition of dissociation under zero friction and applied stress. The condition $F_{int} = F_\gamma$ is recovered and the dissociation width matches with $d_o$ give in equation 7. In summary, under an applied stress in the $d_-$ method (Figure 5), Partial_1 starts to move first in the positive x-direction and the separation between partials increases to $d_o$. In contrast, when the dislocation is introduced using the $d_+$ method (Figure 5), it is Partial_2 which starts in the positive x-direction getting closer to Partial_1 and the final dissociation width decreases to $d_o$. The equilibrium dissociation width in concentrated alloys can be obtained only during dislocation motion. Of course, when the applied stress is removed after the dislocation motion, the dissociation width remains at $d_o$ and the system "forgets" its initial configuration whether it was $d_+$ or $d_-$.

To confirm these theoretical findings, we investigate the dissociation width after relaxation, with the edge dislocation introduced using the two methods ($d_+$ and $d_-$) and compare it with that computed during dislocation motion $d_o$ as a function of temperature in the ASS alloy. The results obtained using EAM-11 potential are shown in Fig.6a.

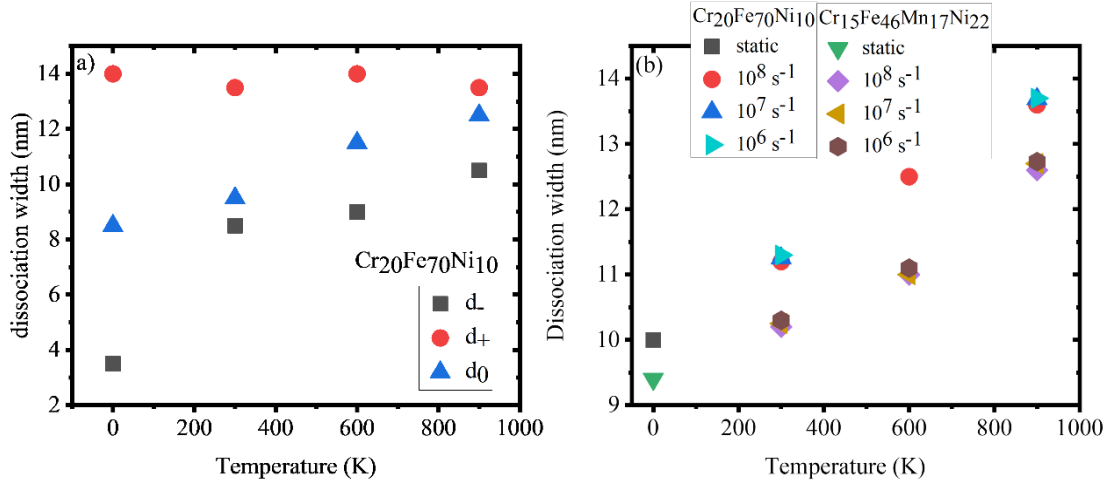

Figure 6: Dissociation width vs temperature. (a) effect of the dislocation introduction method ($d_+$ and $d_-$) on the dissociation width in the relaxed ASS alloy ($Cr_{20}Fe_{70}Ni_{10}$) using the EAM-11 potential compared with the dissociation width measured during motion $d_o$. (b) effect of the strain rate and temperature on the dissociation under the motion of the edge dislocation in the HEA ($Cr_{15}Fe_{46}Mn_{17}Ni_{22}$) and ASS ($Cr_{20}Fe_{70}Ni_{10}$) alloys using the EAM-21 potential

In Fig. 6a, we notice that in agreement with the predictions: (i) $d_+$ remains constant as predicted since $F_f > F_\gamma - F_{int}$ in the whole temperature range; (ii) $d_-$ increases significantly with temperature also as predicted by eq. 8 owing to the large decrease in the friction stress (see next section); (iii) the measured dissociation width during motion $d_o$ is as expected between $d_+$ and $d_-$. It increases slightly with temperature due to the slight decrease in the shear modulus.

Fig. 6b, shows the evolution of the dissociation width during its motion as calculated for both alloys using the EAM-21 potential at different strain rates and temperatures. First, we notice that strain rate does not affect the dissociation width, which slightly increases with temperature owing to the decrease in the shear modulus. Second, the dissociation width in the ASS alloys ($Cr_{20}Fe_{70}Ni_{10}$) remains larger than in the HEA alloy ($Cr_{15}Fe_{46}Mn_{17}Ni_{22}$). This is of course attributed to the fact that the SFE in the ASS alloys is lower than in the HEA alloy.

## 3.4 Dislocation mobility

As a general statement, molecular dynamics results show that the average shear stress required to move the dislocation is higher in the $Cr_{15}Fe_{46}Mn_{17}Ni_{22}$ HEA alloy than in the $Cr_{20}Fe_{70}Ni_{10}$ ASS alloy at all temperatures and strain rates. The motion of the partial dislocations is always jerky. This is evidenced by the large number of peaks on the stress-strain curves, see Fig. 7. Partials are continuously pinned by many local strong atomic configurations. Every stress peak corresponds to an unpinning event.

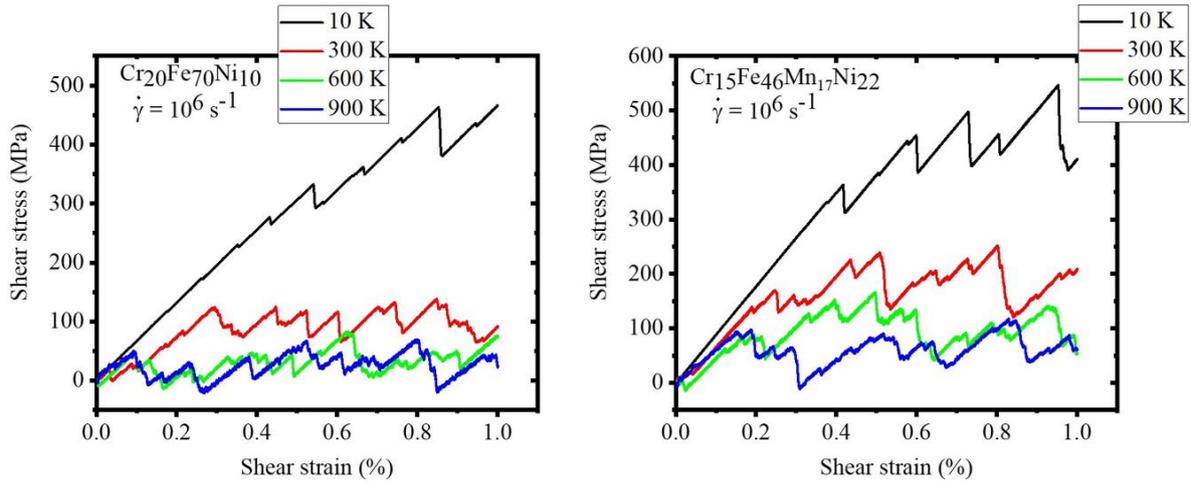

Figure 7: Examples of the loading curves of the simulation box (using EAM-21) containing an edge dislocation at different temperatures of the ASS ($Cr_{20}Fe_{70}Ni_{10}$) alloy and of the HEA ($Cr_{15}Fe_{46}Mn_{17}Ni_{22}$). The black, red, green and blue curves refer to simulations at 10, 300, 600 and 900 K, respectively.

In the following, the critical stress for the dislocation motion at a given stress and strain rate is computed as an average of the stress curve recorded beyond the elastic loading. However, the dislocation velocity is computed from the average dislocation position, taken as the average positions of the partial dislocations. Figure 8 shows the evolution of the critical stress as a function of the dislocation velocity at different simulation temperatures.

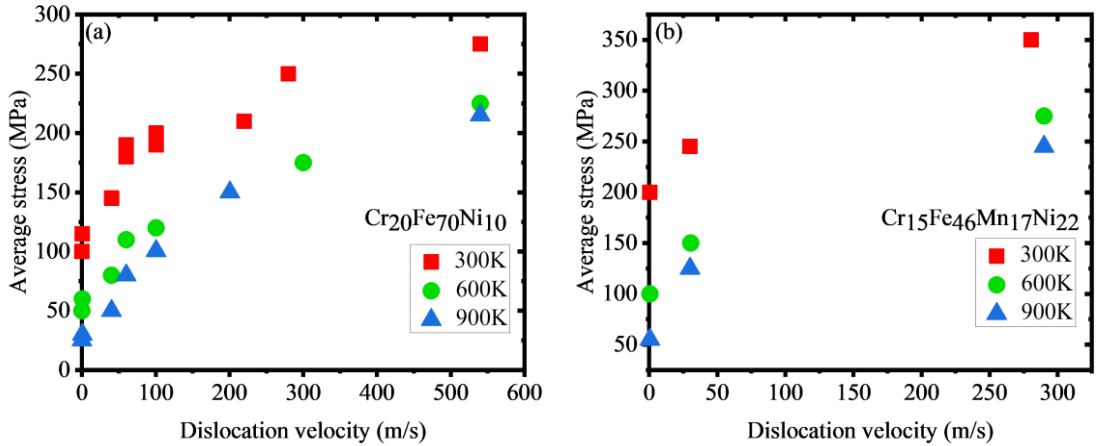

Figure 8: Evolution of the critical resolved shear stress (CRSS) on the (111) plane as a function of the temperature at different strain rates for the (a) ASS alloy ($Cr_{20}Fe_{70}Ni_{10}$) using EAM-11 and EAM-21 and the (b) HEA alloy ($Cr_{15}Fe_{46}Mn_{17}Ni_{22}$) using EAM-21.

The result shows that, as expected from the thermal activation theory, the average stress during dislocation motion decreases strongly with temperature and increases with strain rate in both alloys.

Results obtained with the two potentials (EAM-11 and EAM-21) on the ASS alloy ($Cr_{20}Fe_{70}Ni_{10}$ alloy) using different MD codes and different simulation box dimensions are found to collapse on the same curves in Fig. 8, which confirms the consistency of the two potentials. It can be clearly seen in the figure that the evolution is not linear in both alloys. The tendency to saturation in the critical stress is not due to the limit of sound speed, since the latter is much larger than the velocities shown in the Figure. In the following, a constitutive equation was proposed for the description of behaviour shown in Figure 8.

In order to correctly describe the dislocation behaviour, two regimes must be distinguished. As widely known [65], during dislocation motion in pure materials, dislocation interacts with phonons (friction regime). At velocities much lower than the sound velocity, this resistance is viscous and linear in FCC materials, see for example [66]. The velocity $v_f$ in this regime is well described by a simple friction model:

$$v_f = \frac{\tau b}{B} \quad (9)$$

where B is the drag coefficient. For the sake of simplicity, we consider a constant average value for the drag coefficient $B = 4.5 \cdot 10^{-5}$ Pa.s. This value is consistent with other investigations (ex. [67]). However, in highly concentrated alloys, where some hard local configurations require unpinning by thermal activation, another regime prevails especially at low temperature/low stress (thermal regime). Although many sophisticated models were reported in the literature, it turns out that a simple Arrhenius type equation correctly predicts the behaviour of the dislocation in these simulations. The thermally activated velocity of the dislocation $v_{th}$ is given by:

$$v_{th} = v_o exp - \frac{\Delta H}{kT} \quad (10)$$

where $v_o$ is a constant, $\Delta H$ is the stress-dependent activation enthalpy and $k$ is the Boltzmann constant. In general, dislocations are submitted simultaneously to the friction resistance (eq. 9) and to the thermally activated rate (eq. 10). Consequently, there is no way to completely isolate the friction regime from the thermally activated one. Away to decrease the thermally activated component is to strongly increase the stress, which results in a low activation energy. But in this case, the velocity becomes very high and may start saturating close to shear wave velocity. A superposition rule is thus required. For the dislocation to move, the applied stress must be large enough to overcome the phonon resistance and to provide enough shear rate for thermal activation. As discussed in [68], the two resistances must add in parallel which implies that the final dislocation velocity matches with the harmonic mean of both velocities:

$$\frac{1}{v} = \frac{1}{v_f} + \frac{1}{v_{th}} \quad (11)$$

The value of $v_o = 2000$ m s$^{-1}$ was found to ensure a smooth evolution of $\Delta H$ with the shear stress $\tau$ in all simulations (different temperatures and strain rates). Using data in Fig. 8, the obtained activation enthalpy as a function of stress is shown in Fig 9 for the two alloys.

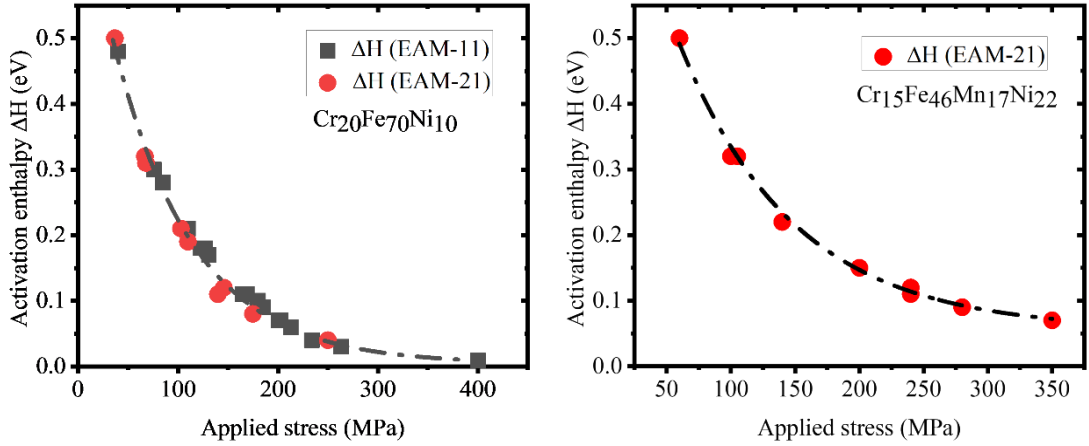

Figure 9: Evolution of the activation enthalpy as a function of the applied shear stress in ASS ($Cr_{20}Fe_{70}Ni_{10}$) and HEA ($Cr_{15}Fe_{46}Mn_{17}Ni_{22}$) alloys.

As expected, the activation enthalpy is a monotonously decreasing function of the stress. Again, results of the two potentials (EAM-11 and EAM-21) on the ASS alloy ($Cr_{20}Fe_{70}Ni_{10}$ alloy) using different MD codes and different simulation conditions are found to collapses on the same curve in Fig. 9. Comparing the results of the two investigated alloys in Fig.9, one notice that activation enthalpy at given stress is larger in the HEA alloy than in the ASS alloy. To provide a constitutive equation for the dislocation motion, one must provide an analytical expression for $\Delta H$ as a function of shear stress $\tau$. Following the framework of Kock et al [69], fitting attempts show that the enthalpy barrier for both alloys can be described by the unique expression:

$$\Delta H = \Delta H_o \left( 1 - \sqrt{\frac{\tau}{\tau_o}} \right)^2 \qquad (12)$$

in which $\Delta H_o$ is the total activation enthalpy (i.e. under zero applied stress) and $\tau_o$ is the theoretical threshold stress at absolute zero temperature. The other coefficients depend on the alloy composition: $\Delta H_o$ = 0.8 eV and $\tau_o$ = 400 MPa for the ASS ($Cr_{20}Fe_{70}Ni_{10}$ alloy) alloy and $\Delta H_o$ = 1.1 eV and $\tau_o$ = 550 MPa for the HEA ($Cr_{15}Fe_{46}Mn_{17}Ni_{22}$) alloy. The higher value of $\tau_o$ in the HEA alloy suggests that the local pinning forces are stronger than in the ASS model alloy. To check the values of $\tau_o$ found in our identification process, the ideal solution is to carry out molecular static simulations (i.e. at 0 K). However, energy minimization is required in molecular statics, which is much more time consuming especially for our large box size. Instead of that, we carried out simulations at 10 K (see Fig. 7). The values found for the average flow stress are lower but quite close to the identified values of $\tau_o$, which validate our superposition rule (the harmonic mean). The comparison between the dislocation velocity computed in MD simulations and the predicted velocity using eq. 11 (together with eq. 9 and 10) are illustrated in Fig. 10a.

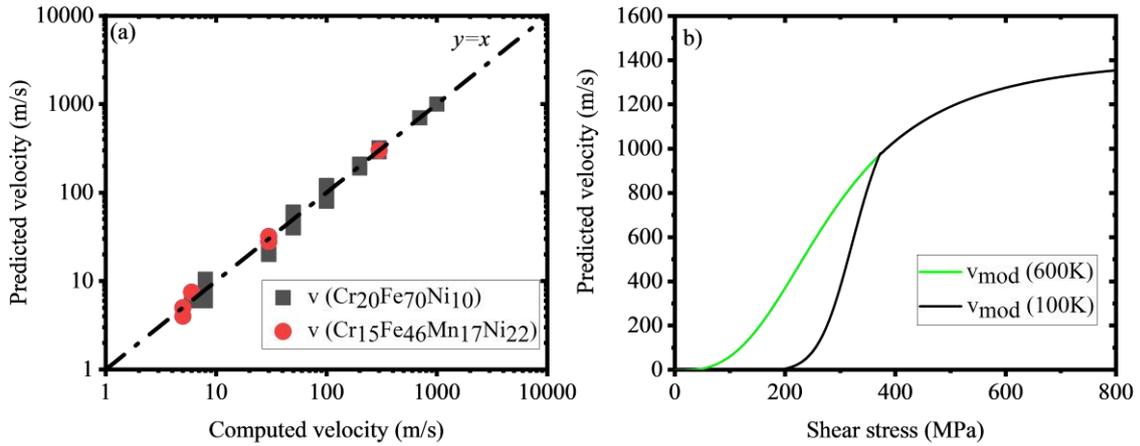

Figure 10: (a) Comparison between the predicted and computed dislocation velocities in all simulations and (b) illustration of the predicted dislocation velocity as a function of shear stress in the ASS alloy ($Cr_{20}Fe_{70}Ni_{10}$) at two different temperatures.

The predicted values of the velocity agree closely with the computed ones over at least three orders of magnitudes. This confirms our approach in superposing the thermally activated regime to the friction regime using the harmonic mean of the velocities. To illustrate the consequences of this superposition rule, we show in Fig. 10b the evolution of the velocity as a function with the applied shear stress at two different temperatures 100 and 600 K. At low stress, the dislocation velocity is strongly dependent on temperature, which is characteristic of the thermally activated regime. However, at high stress, there is no effect of the temperature on the dislocation velocity which indicates that the controlling regime is the friction stress.

## 4 Conclusion

In this study, the elementary dislocation properties in a model alloy of austenitic stainless-steel ASS, ($Cr_{20}Fe_{70}Ni_{10}$) and a model high entropy alloy HEA, ($Cr_{15}Fe_{46}Mn_{17}Ni_{22}$) were investigated using two potentials and different MD codes at different temperatures and strain rates. Based on the above presented and discussed results the following conclusions can be drawn:

I. There is a strong fluctuation of the local SFE in both alloys, that can accurately be described by a Gaussian distribution function, in agreement with experimental observations. The SFE fluctuation is higher in $Cr_{15}Fe_{46}Mn_{17}Ni_{22}$ alloy compared to $Cr_{20}Fe_{70}Ni_{10}$ alloy.

II. The critical stress for the dislocation motion of the ½<110>{111} edge dislocation is larger in the $Cr_{15}Fe_{46}Mn_{17}Ni_{22}$ in comparison with the $Cr_{20}Fe_{70}Ni_{10}$ alloy. The critical stress decreases strongly with temperature and strain rate, which is consistent with experimental investigations of solid solution strengthening.

III. The method used to introduce the dislocation in the simulation box is found to strongly alter the dissociation width obtained after relaxation. A theoretical explanation is presented to explain this issue and to confirm that an equilibrium dissociation width is achieved during dislocation motion. Moreover, the equilibrium dissociation width of $Cr_{20}Fe_{70}Ni_{10}$ alloy is larger than the width of $Cr_{15}Fe_{46}Mn_{17}Ni_{22}$ alloy. The difference can be traced to the SFE values, the SFE of ASS alloys is lower than that of HEA alloy.

IV. The dislocation velocity varies as a function of the applied stress and simulation temperature. The velocity is found to obey a harmonic mean superposition rule involving a viscous regime induced by phonon drag and a thermally activated regime induced by local dislocation pinning due to strong fluctuation in the local chemical composition. The provided constitutive equation is found to accurately predict the dislocation velocity over three orders of magnitudes in the $Cr_{20}Fe_{70}Ni_{10}$ and $Cr_{15}Fe_{46}Mn_{17}Ni_{22}$ alloys. Therefore, the dislocation velocity is found to be faster in ASS when compared to HEA.

# 5 Data availability

All data included in this study are available upon request by contact with the corresponding author.

## Acknowledgements

This work is funded under the French ANR-PRCE-HERIA project (ANR-19-CE08-0012-01). One of the authors (GB) acknowledges funding from the Euratom research and training program 2014–2018 under grant agreement No. 755269 (GEMMA project).

# Appendix A

*Temperature-dependence of the single elastic constant and the shear modulus $\mu$*

There are three methods for estimating the elastic constants: (a) direct approach, which involves calculating the change in internal energy due to applied strain, (b) calculating the box fluctuation with Canonical ensemble; NPH (number of atoms, pressure, and enthalpy are constant), or NPT (number of atoms, pressure, and enthalpy are constant) (number of atoms, pressure, and temperature are constant) and (c) Using microcanonical ensembles and keeping volume constant, the elastic constant can be calculated as the average of stress fluctuations; NVE (number of atoms, volume, and energy are constant) or NVT (number of atoms, volume, and energy are constant) (number of atoms, volume, and temperature are constant).

The third method was employed to maintain the volume of the box when temperature was applied. The simulation box comprises approximately 30000 atom crystals that are subjected to <111> tensile loading with periodic boundary conditions applied in all directions. When the temperature was applied, NPT was employed to maintain equilibrium. After the system has been relaxed, NVT is applied to the MD simulation to achieve system equilibration so that the elastic constant may be calculated with temperature.

The isothermal stress fluctuations formula was utilized to analyze the MD results in order to derive the elastic constant for finite temperature, as follows [70-71]:

$$C_{ijkl} = \langle \chi^B_{ijkl} \rangle - \frac{V}{k_B T} \left( \langle P_{ij} P_{il} \rangle - \langle P_{ij} P_{kl} \rangle \right) + \frac{2 N k_B T}{V} (\delta_{ik} \delta_{jl} + \delta_{il} \delta_{jk}) \quad (A.1)$$

where the first term is the Born term, which is an elastic constant at 0 K and is defined as [47]:

$$\chi^B_{ijkl} = \frac{1}{V} \frac{\partial^2 \langle E \rangle}{\partial \varepsilon_{ij} \partial \varepsilon_{kl}} \quad (A.2)$$

$\langle E \rangle$ is the average energy per atom, while V is the atomic volume, $N$ is the number of atoms and $k_B$ is the Boltzmann constant.

The stress tensor fluctuation is the second term in equation (A.1), and the kinetic energy is the last term.

It is important to note that, the lattice parameter used for the MD calculation are also temperature dependance, the figure A.1 shows the evolution of the equilibrium lattice parameter with temperature for the two study alloys.

The method used follows the approach used in [72].

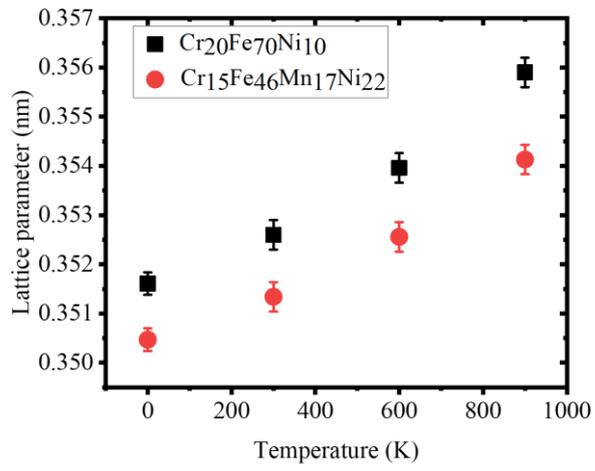

Figure A1: Equilibrium lattice parameter with temperature for HEA ($Cr_{15}Fe_{46}Mn_{17}Ni_{22}$) and ASS ($Cr_{20}Fe_{70}Ni_{10}$) alloys using the EAM-21 potential